# Interface control of the magnetic chirality in TaN|CoFeB|MgO heterosctructures

Jacob Torrejon[1], Junyeon Kim[1], Jaivardhan Sinha[1], Seiji Mitani[1] and Masamitsu Hayashi[1*]

[1]*National Institute for Materials Science, Tsukuba 305-0047, Japan*

Michihiko Yamanouchi[2,3] and Hideo Ohno[2,3,4]

[2] *Center for Spintronics Integrated Systems, Tohoku University, Sendai 980-8577, Japan*

[3]*Research Institute of Electrical Communication, Tohoku University, Sendai 980-8577, Japan*

[4]*WPI Advanced Institute for Materials Research, Tohoku University, Sendai 980-8577, Japan*

**Recent advances in the understanding of spin orbital effects in ultrathin magnetic heterostructures have opened new paradigms to control magnetic moments electrically[1,2]. The Dzyaloshinskii-Moriya interaction[3,4] (DMI) is said to play a key role in forming a Neel-type domain wall that can be driven by the spin Hall torque[5-8], a torque resulting from the spin current generated in a neighboring non-magnetic layer via the spin Hall effect[9-11]. Here we show that the sign of the DMI, which determines the direction to which a domain wall moves with current, can be changed by modifying the adjacent non-magnetic layer. We find that the sense of rotation of a domain wall spiral[12] is reversed when the Ta underlayer is doped with nitrogen in Ta|CoFeB|MgO heterostructures. The spin Hall angle of the Ta and nitrogen doped Ta underlayers carry the same sign, suggesting that the sign of the DMI is defined at the interface. Depending on the sense of rotation, spin transfer torque and spin Hall torque can either compete or assist each other, thus influencing the efficiency of moving domain walls with current.**

*Email: hayashi.masamitsu@nims.go.jp

Understanding the underlying physics of current driven domain wall motion is essential in developing advanced storage class memory devices[13]. Conventionally, domain walls move along the electron flow (against the current) when driven by spin transfer torque[14-16]. Recently, a number of experiments have shown that the domain walls can instead move against the electron flow in magnetic heterostructures[7,8,17-21]. To describe this effect, the spin Hall effect[9-11] in the heavy metal layer has been considered as a possible source of the spin current. The generated spin current diffuses into the ultrathin magnetic layer and exerts torque, termed the "spin Hall torque[2]", on the domain wall magnetization only if the wall forms a Neel-type wall[5]. To move sequences of domain walls with current in the same direction, the Neel wall has to alternate its chirality between neighboring domain walls. This requires formation of a "domain wall spiral[12]", which can be generated in systems with large spin orbit coupling and broken inversion symmetry via the Dzyaloshinskii-Moriya interaction[3,4] (DMI).

In the above model[5-8], the direction to which a domain wall moves with current is determined by the signs of the spin Hall effect and the DMI. The sign of the spin Hall effect depends on the heavy metal layer and is determined by the element specific spin orbit coupling; for example, it is opposite[2,22,23] for Pt and Ta. For the DMI, the sign depends upon the spin orbit coupling as well as the structural symmetry of the magnetic layer[3,4]. For example, in three dimensional bulk-like systems, the sense of rotation of the magnetic structure, i.e. the "chirality", can either follow or be opposite to the crystallographic chirality in Mn and Fe based non-centrosymmetric B20 structures[24,25], respectively, indicating the difference in the spin orbit coupling of the Mn and Fe based systems. The magnetic chirality at surfaces has been studied intensively using spin polarized scanning tunneling microscopy[26,27]. Here the surface atomic configuration plays an important role in setting the chirality.

The origin of the DMI at interfaces is more difficult to treat as the structural symmetry determination is non-trivial. It has been reported that DMI changes its sign depending on the order of the film stack[7,28], which is consistent with the three-site indirect exchange mechanism[29,30] proposed previously. Recent experiments[8] have indicated that for a given magnetic layer (CoFe) the sign of the DMI is the same even when the adjacent non-magnetic layer (Pt or Ta) has opposite sign of the spin orbit coupling constant.

Here we show that the sign of the DMI can be changed for a given magnetic layer when the neighboring non-magnetic layer is modified. In Ta|CoFeB|MgO heterostructures we find that the domain wall moves along the electron flow, whereas it propagates against the electron flow when the Ta layer is doped with nitrogen. The sign of the spin Hall effect is the same for Ta and nitrogen doped Ta, indicating that, surprisingly, the sign of the DMI is reversed between the two systems. The Ta- and TaN-CoFeB interface is predominantly amorphous, thus the sign change is likely related to modification of the spin orbit coupling at the interface. The strength of the DMI is weak compared to recently reported systems with Co|Ni multilayers[7] and we find that both the spin Hall torque and the conventional spin transfer torque can drive the domain wall, which gives rise to efficient wall motion when the two torques work in sync.

Films are deposited using magnetron sputtering on Si(001) substrates coated with 100 nm thick thermally oxidized Si. The film stack is Sub.|d TaN(Q)|1 CoFeB|2 MgO|1 Ta (units in nanometer). TaN is formed by reactively sputtering Ta in a mixture of $N_2$ and Ar gas[31]. Q represents the fraction of $N_2$ gas introduced during the Ta sputtering with respect to the entire ($N_2$+Ar) gas. Q is varied from 0, which corresponds to pure Ta, to 0.7% and 2.5%. The thickness of the Ta(N) layer is varied within in the substrate using a linear shutter during the sputtering. Magnetic easy axis points out of plane owing to the perpendicular magnetic anisotropy developed at the CoFeB|MgO interface[32]. Wires are patterned using photo-

lithography and Ar ion etching followed by a lift-off process to form the electrical contacts made of 10 Ta| 100 Au.

An optical microscopy image of the wire along with schematic illustration of the experimental setup is shown in Fig. 1(a) and 1(b). Variable amplitude voltage pulses (duration fixed to 100 ns) are fed into the wire from a pulse generator. Positive voltage pulse supplies current into the wire that flows along the +X direction. A domain wall is nucleated[33] by applying a voltage pulse above a critical amplitude which depends on the film stack. The out of plane field needed to move a domain wall, i.e. the propagation field, is small for most of the wires: typically below ~20 Oe (see supplementary information for the details). Kerr microscopy is used to acquire magnetic images of the sample and current driven domain wall velocity is estimated by dividing the distance the wall traveled by the pulse length. Typical wires studied here have a width of ~5 μm.

Exemplary Kerr microscopy images are shown in Fig. 1(c) and 1(d) when negative and positive voltage pulses are applied to devices made of Ta and TaN(Q=0.7%) underlayers, respectively. The domain wall moves along the electron flow for the former whereas it moves against the electron flow for the latter. Note that the domain walls shown in Fig. 1(c)-(d) correspond to same configuration (↓↑ walls). Depending on the thickness and dielectric constant of the each layer including the 100 nm thick $SiO_2$, the Kerr contrast can change (see supplementary information).

Domain wall velocity as a function of the voltage pulse amplitude is summarized in Fig. 2 for the three film structures. Positive velocity corresponds to a domain wall moving toward the +X direction. For the Ta underlayer films, the domain wall always moves along the electron flow. This also applies for the TaN(Q=0.7%) underlayer when the underlayer thickness $d$ is less than ~1.6 nm. However, the domain wall moves against the electron flow when $d$ is larger than ~1.6 nm for the TaN(Q=0.7%) underlayer and all film thicknesses studied for the TaN(Q=2.5%) underlayer. The threshold voltage needed to induce domain wall motion decreases as the underlayer thickness $d$ is increased for all film structures. The applicable pulse amplitude is limited by the voltage at which current induced domain wall nucleation occurs. The corresponding current density below which a domain wall moves without inducing nucleation is shown in the supplementary material.

To examine the underlying mechanism of domain walls moving against the electron flow, we study the current induced effective magnetic field in a Hall bar patterned on the same substrate using the adiabatic (low frequency) harmonic Hall voltage measurements[34]. The effective field provides information on the size and sign of the spin Hall torque[35]. Figure 3 shows the effective field components directed transverse to (Fig. 3a-c) and along (longitudinal) (Fig. 3d-f) the current flow direction plotted as a function of the underlayer thickness $d$. Interestingly, the $d$ dependence of the effective field is similar among the film structures studied here: both the transverse and longitudinal components increase in magnitude with increasing $d$ and the direction of each component is the same for all structures when $d$ is large. As we reported previously[35], the source of the effective field that increases with increasing underlayer thickness is likely the spin Hall torque, whose direction is determined by the spin Hall angle. Thus these results show that the sign of the spin Hall angle is the same for all films with different underlayers considered here.

However the magnitude of each component, in particular, the thick underlayer limit which is proportional to the spin Hall angle[2], varies depending on the nitrogen doping concentration (see supplementary information for more details). The change in the effective field may be partly related to the boron concentration in the underlayer which decreases with increasing nitrogen concentration[31] and can influence the scattering rate. With regard to current induced domain wall motion, it is the longitudinal component that drives a Neel wall[5,6]: the transverse component influences domain wall nucleation[33]. Note that the effective field changes its sign at a critical thickness $d_C$ for all films. $d_C$ increases with increasing nitrogen concentration which illustrates a competition between the spin Hall torque and a Rashba-like interface torque in each film[35]. Since the size of the latter torque is small, we do not take this into account in describing the motion of domain walls driven by current.

As the sign of the spin Hall angle is the same for all film structures, we infer that the DMI is changing its sign between the Ta and the nitrogen doped Ta underlayers. To confirm this hypothesis, we study the in-plane magnetic field dependence of domain wall velocity. In out of plane magnetized systems, the preferred domain wall configuration is the Bloch type for the wire dimension used here: a Neel wall is only stable for narrow wires (typically below ~100 nm) where shape anisotropy starts to dominate[16,36]. However, the DMI can promote a Neel type wall with a fixed chirality. This interaction can be modeled as an additional offset field applied along the wire's long axis for a given domain wall[5,7,8]. The offset field changes its direction depending on whether the magnetization of the neighboring domain points ↑↓ or ↓↑, thus forming a domain wall spiral[12]. An in-plane field directed along the wire's long axis ($H_L$) can thereby effectively modify the offset field and influence the wall velocity[7,8].

Figure 4 shows representative results for two different structures in which the domain wall moves in opposite directions when driven by current. The velocity scales almost linearly with $H_L$ in all cases. At zero $H_L$, both ↑↓ or ↓↑ walls move in the same direction for a given film structure. However, the slope and consequently the field at which the velocity becomes zero (defined as $H_L^*$ hereafter) is different depending on the film structure and the wall type. For example, $H_L^*$ is positive (negative) for a ↑↓ (↓↑) wall when the wall moves along the electron flow (Fig. 4a). This indicates that there is a negative (positive) "offset field" associated with the ↑↓ (↓↑) wall. The direction of this offset field ($H_L^*$) reverses for systems with wall moving against the electron flow (Fig. 4b). These results show that the domain wall spiral possess a left handed chirality (↑←↓ and ↓→↑: the colored arrow indicating the magnetization direction of the Neel wall) for the walls moving along the electron flow (Ta underlayer) and it forms a right handed chirality (↑→↓ and ↓←↑) when the direction of the wall motion reverses (TaN underlayer).

The underlayer thickness dependence of $H_L^*$ is plotted in Fig. 5(a-c). The slope of the velocity vs. $H_L$ ($v_{DW}/H_L$) is shown in Fig. 5(d-f). According to the one-dimensional (1D) model of a domain wall, $H_L^*$ scales with the DMI effective field, which is proportional to the strength of DMI and the inverse of the domain wall width[5,7,8]. In Fig. 5(a-c), we find a clear correlation between the direction of the wall motion and the sign of $H_L^*$, suggesting that the DMI is changing its sign with the underlayer material. The variation in the size of $H_L^*$ is partly related to change in the magnetic anisotropy ($K_{EFF}$), which changes the domain wall width and thereby influencing $H_L^*$ ($H_L^*$ scales with $K_{EFF}^{1/2}$) [5,7]. $K_{EFF}$ decreases rapidly with the Ta underlayer thickness whereas it is more or less constant with the underlayer thickness for the TaN underlayer films[31] (see supplementary information). This trend is also evident in the underlayer thickness dependence of $v_{DW}/H_L$ (Fig. 5(d-f)), which, in contrast to $H_L^*$, increases with the inverse of $K_{EFF}^{1/2}$.

The size of $H_L^*$ is also dependent on the relative magnitude of the spin Hall and spin transfer torques (see supplementary information for the details). In the 1D model, spin transfer torque can also influence $H_L^*$ as long as a non-zero spin Hall torque is present. We infer that the sign reversal of $H_L^*$ in Fig. 5(b) for the TaN(Q=0.7%) underlayer at $d$~1.6 nm is related to the competition between the two torques. Since the current density flowing through the CoFeB layer is large for films with small TaN(Q=0.7%) underlayer thickness (see supplementary information), spin transfer torque dominates the wall motion in this regime. We conjecture that the sign of DMI does not vary with the TaN(Q=0.7%) underlayer thickness, i.e. the walls are all right handed. Similarly, for the Ta underlayer films, one may consider that the wall is right handed, in contrast to what have been reported recently[8], and the spin transfer torque dominates. However, since the Ta underlayer is less resistive than the TaN underlayers, less current flows through the CoFeB layer and thus we consider that the spin transfer torque is weaker in this system. Moreover, the velocity tends to increase when the Ta underlayer thickness is increased for a given voltage above the threshold (Fig. 2(a)), thus indicating that both torques (spin Hall and spin transfer) are working in sync and thereby the wall is left handed.

Since both Ta|CoFeB and TaN|CoFeB interfaces are structurally similar (amorphous), it is difficult to associate the respective change in the wall chirality, from left to right handed walls, with the change in the structure inversion asymmetry (although there remains a possibility that a short range ordering, which is difficult to identify from transmission electron microscopy images, contributes to the asymmetry). Based on a three-site indirect exchange mechanism[29,30], such change in the wall chirality, i.e. the change in the DMI sign, may be related to the difference in the position of the strong spin orbit coupling source, whether it is placed above or below the Ta(N)|CoFeB interface. Since the boron concentration in the CoFeB layer increases and the magnetic dead layer thickness decreases by doping the Ta underlayer with nitrogen[31], there are elements, such as Ta, B and N, that change their position with respect to the Ta(N)|CoFeB interface upon nitrogen doping. Alternatively, the DMI sign change may be related to the change in the charge localization of the interface atoms, which has been reported to change the sign of the Rashba spin splitting at metal alloy surfaces[38]. Since nitrogen is known for its large electronegativity, addition of nitrogen may contribute to the sign change of the interface spin orbit coupling. Although the magnitude of the DMI is small in Ta(N)|CoFeB|MgO heterostructures, its sign has a significant influence on the domain wall dynamics as it can change the direction to which a domain wall propagates with current.

**Acknowledgements**

This work was partly supported by the FIRST program from JSPS and the Grant-in-Aid (25706017) from MEXT.

**Author Contributions**

M.H. planned the study. J.T. and M.H. wrote the manuscript. J.S. performed film deposition and film characterization, J.T., J.S. and M.H. fabricated the devices. J.K. evaluated the current induced effective field, J.T. carried out the Kerr measurements and analyzed the data with the help of M.H., M.Y., S.M. and H.O. All authors discussed the data and commented on the manuscript.

**Competing financial interests**

The authors declare that they have no competing financial interests.

**References**

1. Miron, I. M., Garello, K., Gaudin, G., Zermatten, P. J., Costache, M. V., Auffret, S., Bandiera, S., Rodmacq, B., Schuhl, A. & Gambardella, P. Perpendicular Switching of a Single Ferromagnetic Layer Induced by in-Plane Current Injection. *Nature* **476**, 189 (2011).
2. Liu, L., Pai, C.-F., Li, Y., Tseng, H. W., Ralph, D. C. & Buhrman, R. A. Spin-Torque Switching with the Giant Spin Hall Effect of Tantalum. *Science* **336**, 555 (2012).
3. Dzyaloshinskii, I. E. Thermodynamic Theory of Weak Ferromagnetism in Antiferromagnetic Substances. *Sov. Phys. JETP* **5**, 1259 (1957).
4. Moriya, T. Anisotropic Superexchange Interaction and Weak Ferromagnetism. *Phys. Rev.* **120**, 91 (1960).
5. Thiaville, A., Rohart, S., Jue, E., Cros, V. & Fert, A. Dynamics of Dzyaloshinskii Domain Walls in Ultrathin Magnetic Films. *Europhys. Lett.* **100**, 57002 (2012).
6. Khvalkovskiy, A. V., Cros, V., Apalkov, D., Nikitin, V., Krounbi, M., Zvezdin, K. A., Anane, A., Grollier, J. & Fert, A. Matching Domain-Wall Configuration and Spin-Orbit Torques for Efficient Domain-Wall Motion. *Phys. Rev. B* **87**, 020402 (2013).
7. Ryu, K.-S., Thomas, L., Yang, S.-H. & Parkin, S. Chiral Spin Torque at Magnetic Domain Walls. *Nat. Nanotechnol.* **8**, 527 (2013).
8. Emori, S., Bauer, U., Ahn, S.-M., Martinez, E. & Beach, G. S. D. Current-Driven Dynamics of Chiral Ferromagnetic Domain Walls. *Nat Mater* **12**, 611 (2013).
9. Hirsch, J. E. Spin Hall Effect. *Phys. Rev. Lett.* **83**, 1834 (1999).
10. Kato, Y. K., Myers, R. C., Gossard, A. C. & Awschalom, D. D. Observation of the Spin Hall Effect in Semiconductors. *Science* **306**, 1910 (2004).
11. Wunderlich, J., Kaestner, B., Sinova, J. & Jungwirth, T. Experimental Observation of the Spin-Hall Effect in a Two-Dimensional Spin-Orbit Coupled Semiconductor System. *Phys. Rev. Lett.* **94**, 047204 (2005).
12. Heide, M., Bihlmayer, G. & Blugel, S. Dzyaloshinskii-Moriya Interaction Accounting for the Orientation of Magnetic Domains in Ultrathin Films: Fe/W(110). *Phys. Rev. B* **78**, 140403 (2008).
13. Parkin, S. S. P., Hayashi, M. & Thomas, L. Magnetic Domain-Wall Racetrack Memory. *Science* **320**, 190 (2008).
14. Yamanouchi, M., Chiba, D., Matsukura, F., Dietl, T. & Ohno, H. Velocity of Domain-Wall Motion Induced by Electrical Current in the Ferromagnetic Semiconductor (Ga,Mn)As. *Phys. Rev. Lett.* **96**, 096601 (2006).
15. Hayashi, M., Thomas, L., Moriya, R., Rettner, C. & Parkin, S. S. P. Current-Controlled Magnetic Domain-Wall Nanowire Shift Register. *Science* **320**, 209 (2008).
16. Koyama, T., Chiba, D., Ueda, K., Kondou, K., Tanigawa, H., Fukami, S., Suzuki, T., Ohshima, N., Ishiwata, N., Nakatani, Y., Kobayashi, K. & Ono, T. Observation of the Intrinsic Pinning of a Magnetic Domain Wall in a Ferromagnetic Nanowire. *Nature Mater.* **10**, 194 (2011).
17. Kim, K.-J., Lee, J.-C., Yun, S.-J., Gim, G.-H., Lee, K.-S., Choe, S.-B. & Shin, K.-H. Electric Control of Multiple Domain Walls in Pt/Co/Pt Nanotracks with Perpendicular Magnetic Anisotropy. *Appl. Phys. Express* **3**, 083001 (2010).
18. Moore, T. A., Miron, I. M., Gaudin, G., Serret, G., Auffret, S., Rodmacq, B., Schuhl, A., Pizzini, S., Vogel, J. & Bonfim, M. High Domain Wall Velocities Induced by Current in Ultrathin Pt/Co/Alox Wires with Perpendicular Magnetic Anisotropy. *Appl. Phys. Lett.* **93**, 262504 (2008).
19. Miron, I. M., Moore, T., Szambolics, H., Buda-Prejbeanu, L. D., Auffret, S., Rodmacq, B., Pizzini, S., Vogel, J., Bonfim, M., Schuhl, A. & Gaudin, G. Fast

Current-Induced Domain-Wall Motion Controlled by the Rashba Effect. *Nat. Mater.* **10**, 419 (2011).

20  Haazen, P. P. J., Mure, E., Franken, J. H., Lavrijsen, R., Swagten, H. J. M. & Koopmans, B. Domain Wall Depinning Governed by the Spin Hall Effect. *Nat. Mater.* **12**, 299 (2013).

21  Koyama, T., Hata, H., Kim, K. J., Moriyama, T., Tanigawa, H., Suzuki, T., Nakatani, Y., Chiba, D. & Ono, T. Current-Induced Magnetic Domain Wall Motion in a Co/Ni Nanowire with Structural Inversion Asymmetry. *Appl. Phys. Express* **6**, 033001 (2013).

22  Morota, M., Niimi, Y., Ohnishi, K., Wei, D. H., Tanaka, T., Kontani, H., Kimura, T. & Otani, Y. Indication of Intrinsic Spin Hall Effect in 4d and 5d Transition Metals. *Phys. Rev. B* **83**, 174405 (2011).

23  Liu, L. Q., Moriyama, T., Ralph, D. C. & Buhrman, R. A. Spin-Torque Ferromagnetic Resonance Induced by the Spin Hall Effect. *Phys. Rev. Lett.* **106**, 036601 (2011).

24  Grigoriev, S. V., Chernyshov, D., Dyadkin, V. A., Dmitriev, V., Maleyev, S. V., Moskvin, E. V., Menzel, D., Schoenes, J. & Eckerlebe, H. Crystal Handedness and Spin Helix Chirality in Fe1-Xcoxsi. *Phys. Rev. Lett.* **102**, 037204 (2009).

25  Grigoriev, S. V., Potapova, N. M., Siegfried, S. A., Dyadkin, V. A., Moskvin, E. V., Dmitriev, V., Menzel, D., Dewhurst, C. D., Chernyshov, D., Sadykov, R. A., Fomicheva, L. N. & Tsvyashchenko, A. V. Chiral Properties of Structure and Magnetism in Mn_{1-X}Fe_{X}Ge Compounds: When the Left and the Right Are Fighting, Who Wins? *Phys. Rev. Lett.* **110**, 207201 (2013).

26  Bode, M., Heide, M., von Bergmann, K., Ferriani, P., Heinze, S., Bihlmayer, G., Kubetzka, A., Pietzsch, O., Blugel, S. & Wiesendanger, R. Chiral Magnetic Order at Surfaces Driven by Inversion Asymmetry. *Nature* **447**, 190 (2007).

27  Heinze, S., von Bergmann, K., Menzel, M., Brede, J., Kubetzka, A., Wiesendanger, R., Bihlmayer, G. & Blugel, S. Spontaneous Atomic-Scale Magnetic Skyrmion Lattice in Two Dimensions. *Nat. Phys.* **7**, 713 (2011).

28  Chen, G., Zhu, J., Quesada, A., Li, J., N'Diaye, A. T., Huo, Y., Ma, T. P., Chen, Y., Kwon, H. Y., Won, C., Qiu, Z. Q., Schmid, A. K. & Wu, Y. Z. Novel Chiral Magnetic Domain Wall Structure in Fe/Ni/Cu(001) Films. *Phys. Rev. Lett.* **110**, 177204 (2013).

29  Fert, A. & Levy, P. M. Role of Anisotropic Exchange Interactions in Determining the Properties of Spin-Glasses. *Phys. Rev. Lett.* **44**, 1538 (1980).

30  Fert, A., Cros, V. & Sampaio, J. Skyrmions on the Track. *Nat. Nanotechnol.* **8**, 152 (2013).

31  Sinha, J., Hayashi, M., Kellock, A. J., Fukami, S., Yamanouchi, M., Sato, M., Ikeda, S., Mitani, S., Yang, S. H., Parkin, S. S. P. & Ohno, H. Enhanced Interface Perpendicular Magnetic Anisotropy in Ta|Cofeb|Mgo Using Nitrogen Doped Ta Underlayers. *Appl. Phys. Lett.* **102**, 242405 (2013).

32  Ikeda, S., Miura, K., Yamamoto, H., Mizunuma, K., Gan, H. D., Endo, M., Kanai, S., Hayakawa, J., Matsukura, F. & Ohno, H. A Perpendicular-Anisotropy Cofeb-Mgo Magnetic Tunnel Junction. *Nat. Mater.* **9**, 721 (2010).

33  Miron, I. M., Gaudin, G., Auffret, S., Rodmacq, B., Schuhl, A., Pizzini, S., Vogel, J. & Gambardella, P. Current-Driven Spin Torque Induced by the Rashba Effect in a Ferromagnetic Metal Layer. *Nat. Mater.* **9**, 230 (2010).

34  Pi, U. H., Kim, K. W., Bae, J. Y., Lee, S. C., Cho, Y. J., Kim, K. S. & Seo, S. Tilting of the Spin Orientation Induced by Rashba Effect in Ferromagnetic Metal Layer. *Appl. Phys. Lett.* **97**, 162507 (2010).

35  Kim, J., Sinha, J., Hayashi, M., Yamanouchi, M., Fukami, S., Suzuki, T., Mitani, S. &

| | Ohno, H. Layer Thickness Dependence of the Current Induced Effective Field Vector in Ta|Cofeb|Mgo. *Nat. Mater.* **12**, 240 (2013). |
|---|---|
| 36 | Jung, S. W., Kim, W., Lee, T. D., Lee, K. J. & Lee, H. W. Current-Induced Domain Wall Motion in a Nanowire with Perpendicular Magnetic Anisotropy. *Appl. Phys. Lett.* **92**, 202508 (2008). |
| 37 | Malozemoff, A. P. & Slonczewski, J. C. *Magnetic Domain Walls in Bubble Material*. (Academic Press, 1979). |
| 38 | Bentmann, H., Kuzumaki, T., Bihlmayer, G., Blugel, S., Chulkov, E. V., Reinert, F. & Sakamoto, K. Spin Orientation and Sign of the Rashba Splitting in Bi/Cu(111). *Phys. Rev. B* **84**, 115426 (2011). |

**Figure captions**

**Figure 1. Schematic of the experimental setup and magneto-optical Kerr images illustrating current induced domain wall motion**. (a) Optical microscope image of the wire used to study current induced domain wall motion. The Ta|Au electrodes are indicated by the yellow colored region. A pulse generator is connected to one of the Ta|Au electrodes, as schematically shown. (b) Illustration of the experimental setup. The thick black arrows indicate the magnetization of the CoFeB layer. (c,d) Typical Kerr images showing current induced domain wall motion along (c) and against (d) the electron flow for wires with different underlayers: (c) ~0.5 nm thick Ta underlayer, (d) ~3.6 nm thick TaN(Q=0.7%) underlayer. Domain walls in (c,d) are both ↓↑-type. The Kerr contrast changes depending on the layer stack (see supplementary information). Between images: ~-40 V, 100 ns long pulses are applied 12 times for (c) and ~28 V, 100 ns long pulses are applied 20 times for (d).

**Figure 2. Pulse amplitude dependence of domain wall velocity**. (a-c) Domain wall velocity as a function of pulse amplitude plotted for various underlayer thicknesses. The films are composed of (a) Ta, (b) TaN(Q=0.7%) and (c) TaN(Q=2.5%) underlayers. The direction to which the wall moves is indicated in the corners of each panel in (c).

**Figure 3. Current induced effective field vs. the underlayer thickness**. (a) Transverse (a-c) and longitudinal (d-f) components of the current induced effective field are plotted as a function of the underlayer thickness for film stacks with different underlayers: (a) Ta, (b) TaN(Q=0.7%) and (c) TaN(Q=2.5%). The effective field is normalized by the amplitude of the excitation voltage ($V_{IN}$), which scales with the current density flowing through the film. The solid and open symbols correspond to the effective field when the magnetization of the CoFeB layer is pointing along +Z and –Z, respectively.

**Figure 4. Longitudinal field dependence of the current driven domain wall velocity**. Domain wall velocity plotted as a function of the longitudinal field (directed along the current flow and the wire's long axis) for two different film stacks: (a) ~0.5 nm thick Ta underlayer and (b) ~3.6 nm thick TaN(Q=0.7%) underlayer. Blue circles and red triangles indicate the wall velocity when positive and negative voltage pulses are applied, respectively. Left (right) panel shows results for ↓↑ (↑↓) wall. Solid lines are linear fits to the data to obtain $H_L^*$ and $v_{DW}/H_L$. The pulse amplitude is ~±40 V for (a) and ~±28 V for (b).

**Figure 5. Offset field and the slope of velocity vs. longitudinal field illustrating the effect of underlayer on the Dzyaloshinskii-Moriya interaction**. (a-c) The offset field $H_L^*$, i.e. the longitudinal field ($H_L$) at which the velocity becomes zero, and (d-f) slope of $v_{DW}$ versus $H_L$ ($v_{DW}/H_L$) is plotted as a function of underlayer thickness for film stacks with different underlayers: (a,d) Ta, (b,e) TaN(Q=0.7%) and (c,f) TaN(Q=2.5%). Solid and open symbols represent ↑↓ and ↓↑ domain walls, respectively. $H_L^*$ and $v_{DW}/H_L$ are evaluated when the wall is driven either by positive or negative voltage pulses: here, both results are shown together. The background color of each panel indicates the direction to which a corresponding domain wall moves; red: against the electron flow, blue: along the electron flow.

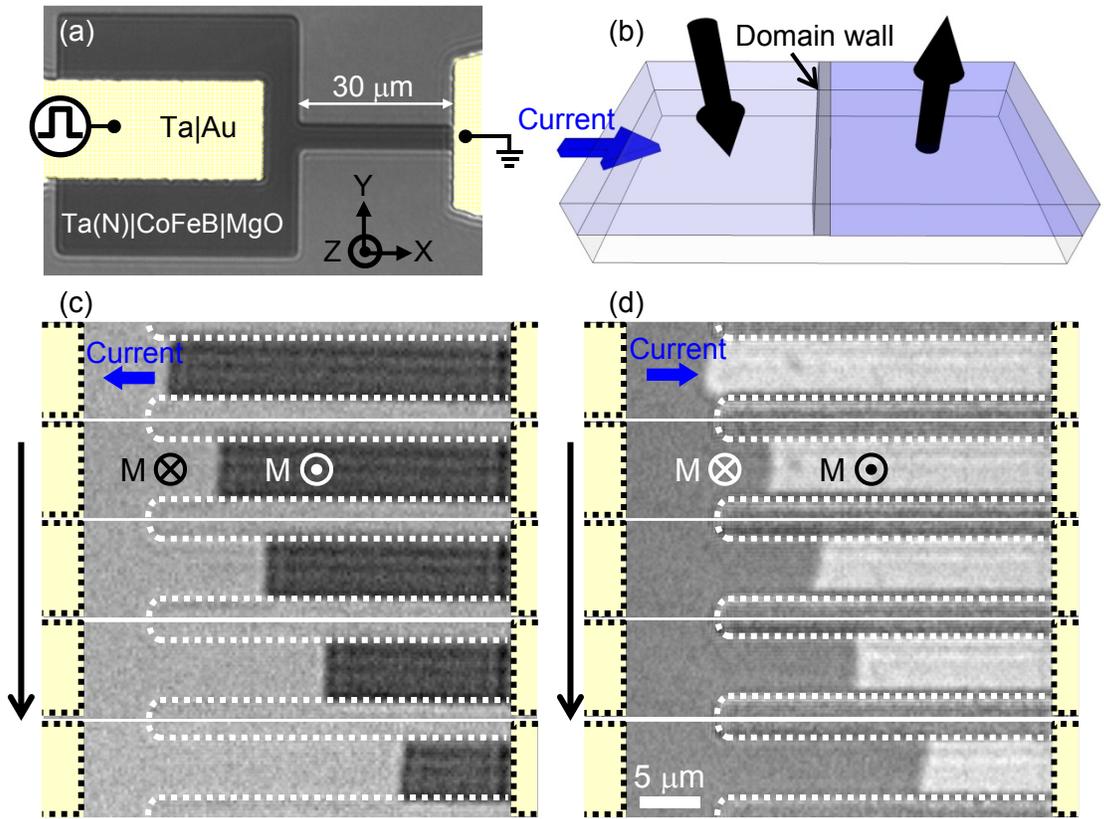

Fig. 1

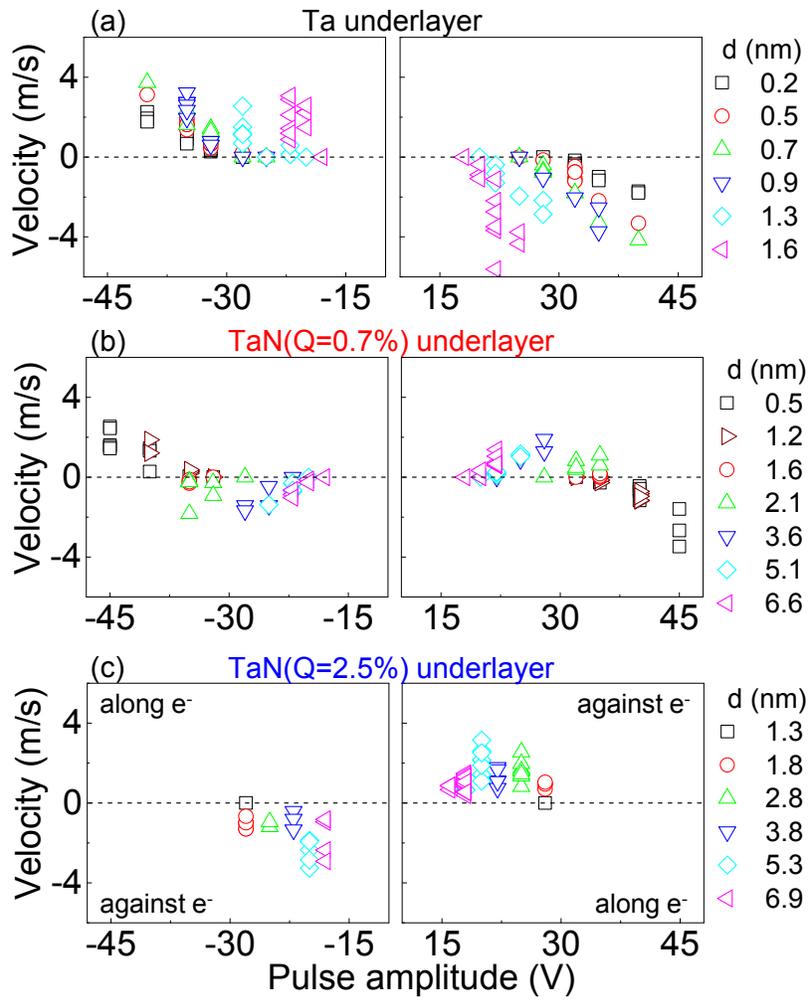

Fig. 2

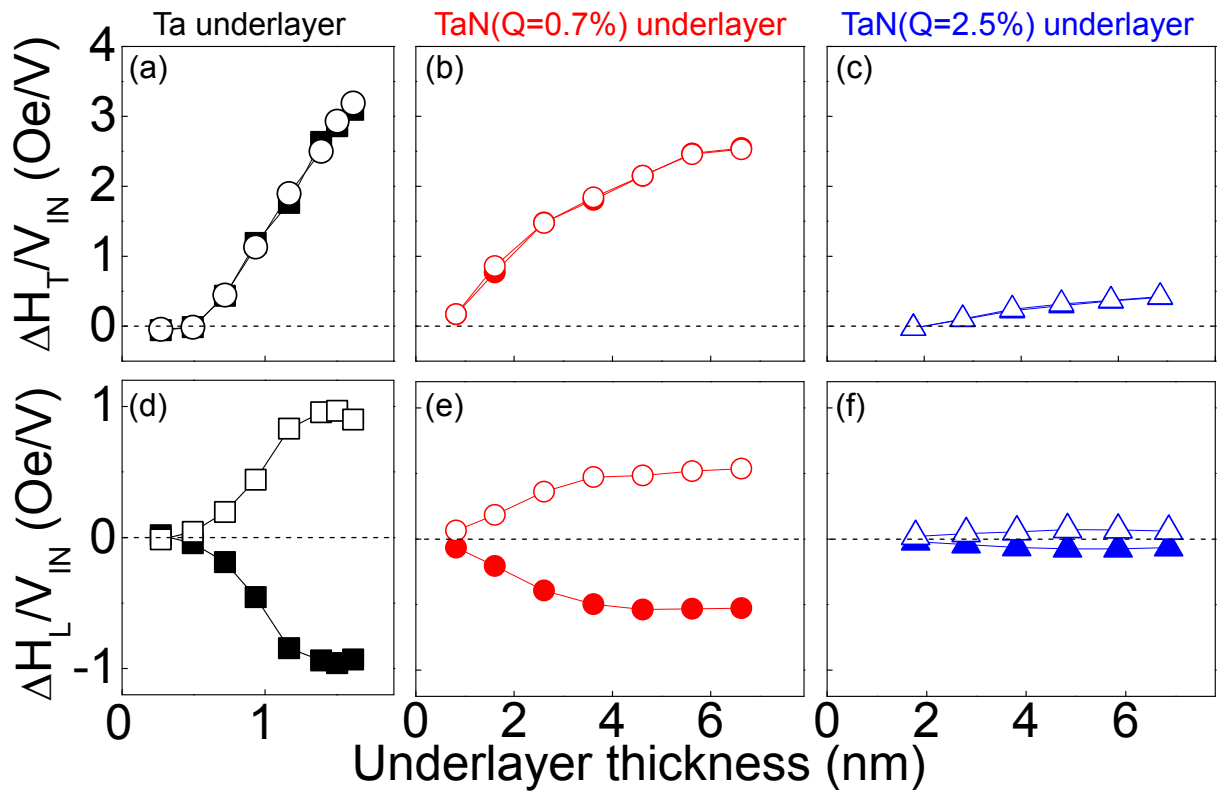

Fig. 3

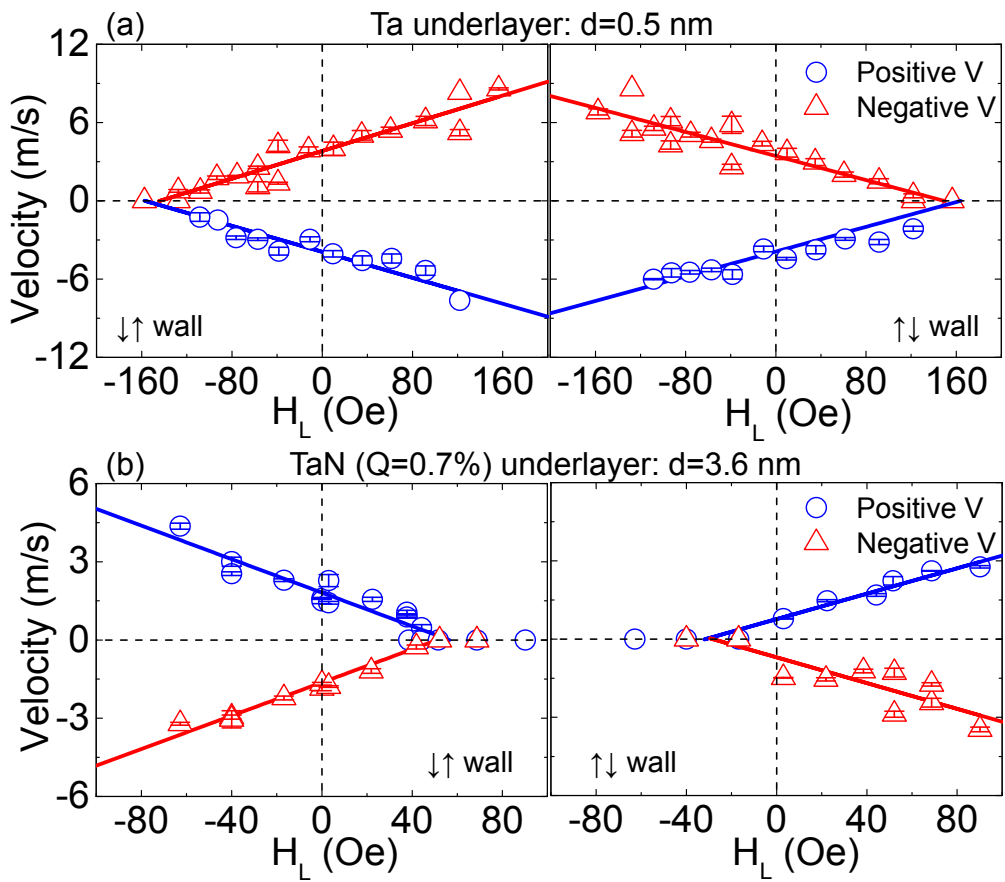

Fig. 4

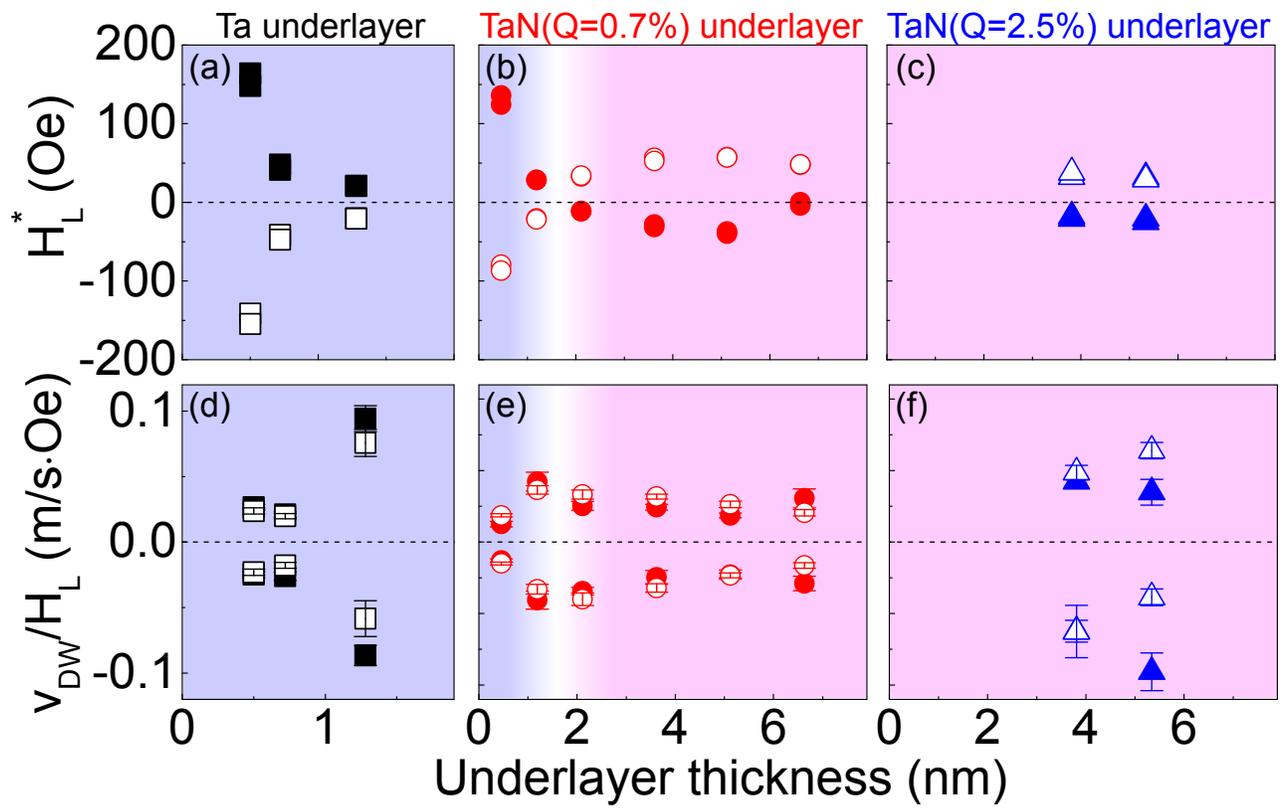

Fig. 5

**Supplementary information for:**

**Interface control of the magnetic chirality in TaN|CoFeB|MgO heterosctructures**

Jacob Torrejon[1], Junyeon Kim[1], Jaivardhan Sinha[1], Seiji Mitani[1] and Masamitsu Hayashi[1*]

[1]*National Institute for Materials Science, Tsukuba 305-0047, Japan*

Michihiko Yamanouchi[2,3] and Hideo Ohno[2,3,4]

[2] *Center for Spintronics Integrated Systems, Tohoku University, Sendai 980-8577, Japan*

[3]*Research Institute of Electrical Communication, Tohoku University, Sendai 980-8577, Japan*

[4]*WPI Advanced Institute for Materials Research, Tohoku University, Sendai 980-8577, Japan*

**S1. Sample preparation**

Films are deposited by magnetron sputtering (DC and RF) on Si|100 SiO$_2$ wafers. The film stack is composed of Substrate|d Ta(N)|1 Co$_{20}$Fe$_{60}$B$_{20}$|2 MgO|1 Ta (units in nanometer). The TaN underlayer is formed by reactively sputtering Ta in the Ar gas atmosphere mixed with a small amount of N$_2$. Ar and N$_2$ gas concentrations are controlled independently by gas mass flow meters. We define Q as the atomic ratio of the N$_2$ gas over the total (Ar + N$_2$) gas, i.e.

$$Q \equiv \frac{S_{N_2}}{S_{Ar} + S_{N_2}},$$ where $S_X$ denotes the mass flow (in unit of sccm) of gas X. $Q$ is varied from 0 to 2.5% here. The resulting atomic composition of TaN is: Ta$_{48}$N$_{52}$ for $Q$=0.7% and Ta$_{44}$N$_{56}$ for $Q$=2.5%, which are determined by Rutherford backscattering spectroscopy (RBS). Note that the atomic composition evaluated by RBS has an error of ±5%. Films are deposited using a linear shutter to vary the underlayer thickness across the wafer. All films are post-annealed at 300 °C

for one hour in vacuum. As shown in Ref. [1], both the Ta underlayer and the CoFeB layers are predominantly amorphous.

Magnetic properties of the films are measured using vibrating sample magnetometry (VSM). The moment per unit volume ($M/V$), anisotropy field ($H_K$) and the magnetic anisotropy energy ($K_{EFF}$) are plotted in Fig. S1. Saturated moment values ($M$) are divided by the product of film area and the CoFeB thickness to obtain $M/V$. It should be noted that the thickness of the CoFeB layer contains, if any, the thickness of a magnetically dead layer. Anisotropy field is defined as the field at which the hard axis hysteresis loop saturates. The magnetic anisotropy energy is estimated from the integrated difference of the out of plane and in-plane hysteresis loops. Positive $K_{EFF}$ corresponds to magnetic easy axis pointing along the film normal. For details, see Ref. [2].

Wires (for evaluating current induced domain wall motion) and Hall bars (for the analysis of current induced effective fields) are patterned by optical lithography and Ar ion etching. Subsequent lift-off process is used to make the 10 Ta|100 Au (units in nm) electrodes.

## S2. Magneto-optical Kerr effect and the hysteresis loops

Magnetic images of the wires are acquired using a Kerr microscopy. To quantify the magnetic contrast, the region of interest (ROI, i.e. the wire) is selected in the acquired image and converted into a two dimensional arrays of integer. The average value of the Kerr intensity (i.e. the CCD signal) of the ROI, denoted as $I$ hereafter, is plotted in Fig. S2(a) as a function of the out of plane field $H_Z$. Hysteresis loops of wires with different TaN(Q=0.7%) underlayer thicknesses are shown. For the thicker underlayer films $I$ is larger when the magnetization is pointing up (large positive $H_Z$). In contrast, $I$ is larger for magnetization pointing downward for the thinner underlayer films. The difference in $I$ when the magnetization is pointing "up" and "down" is defined as $\Delta I$ and the mean value of $I$ is denoted as $I_0$. Figure S2(b) shows $\Delta I/I_0$ as a function of the underlayer thickness for the three film structures investigated here. The sign of $\Delta I/I_0$ changes at a certain thickness for each film structure.

These changes in $\Delta I/I_0$ are likely due to an optical interference effect within the sample. As the total thickness of the heterostructure is very thin, a significant amount of light passes through the heterostructure (Ta(N)|CoFeB|MgO|Ta) and reaches the Si|SiO$_2$ interface (the thickness of the SiO$_2$ is ~100 nm). Magneto-optical Faraday effect takes place when the light transmits through the heterostructure, whereas the Kerr effect contributes to the signal reflected at the film surface. Most of the light which have transmitted through the film reach the Si|SiO$_2$ interface and get reflected to travel toward the heterostructure. Some fraction of the reflected light can transmit through the heterostructure (and again developing the Faraday effect) and propagate toward the CCD camera; the other fraction will get reflected at the heterostructure and again travel toward the Si|SiO$_2$ interface. This will develop interference in the 100 nm thick SiO$_2$ layer and the magneto-optical signal captured with the CCD camera likely includes contribution from both the Kerr and the Faraday effects. Such multiple reflections/interference can change the size and sign of $\Delta I/I_0$. Note that we do not observe any change in the sign of $\Delta I/I_0$ when we use naturally oxidized Si subtrates (with just a few nanometer thick SiO$_2$ layer), confirming that the effect is optical (and not electronic).

All images shown in this paper are subtracted images. An image of a uniformly magnetized state with magnetization pointing along –Z is captured as the reference image. This reference image is subtracted from each image.

## S3. Domain wall nucleation using current pulses

A domain wall is nucleated by applying voltage pulses to the wire. Firstly, the CoFeB layer is uniformly magnetized by applying an out of plane field $H_z$. The field is reduced to zero and we apply a voltage pulse (typically 100 ns of duration) to nucleate a domain wall. This process typically suffices to create one or two domain walls within the wire. In some film structures, an additional out of plane field application is required to change the domain pattern after the pulse application to form an appropriate domain structure.

## S4. Propagation field of the domain walls

The out of plane field needed to move a domain wall, i.e. the propagation field, is evaluated using Kerr microscopy images. After the domain wall nucleation process, the out of plane field $H_Z$ is ramped towards higher magnitude, either to positive or negative $H_Z$, and the magnetic state is monitored with the Kerr microscopy. Such measurement cycle is repeated in each device 10 times (5 times for positive and 5 times for negative $H_Z$). Figure S3(a,b) show typical domain wall motion when $H_Z$ is varied: (a) positive and (b) negative $H_Z$. The propagation field $H_P$ is defined as the field ($H_Z$) at which the Kerr signal change exceeds 50% of the total change expected. The field ramp rate is approximately 1 Oe/sec near the propagation field.

$H_P$ as a function of the underlayer thickness is plotted in Fig S3(c-e) for the three film structures. The change in the $H_P$ is more or less correlated to the magnetic anisotropy of the films: films with larger perpendicular magnetic anisotropy $K_{EFF}$ display larger $H_P$. The linear increase of $H_P$ with increasing underlayer thickness for the TaN(Q=0.7%) underlayer films (Fig. S3(d)) does not exactly follow the trend of the magnetic anisotropy and its origin remains to be identified.

## S5. Estimation of the current density required for domain wall motion

The current density $J$ corresponding to the pulse amplitude needed to move a domain wall, i.e. the maximum pulse amplitude used for each wire as shown in Fig. 2, is plotted in Fig. S4. The solid symbols correspond to $J$ if one assumes a uniform current flow across the Ta(N) and CoFeB layers, defined as $J_{Total}$ hereafter. The open symbols show $J$ for the current that flows through the CoFeB layer (defined as $J_{CoFeB}$) when the resistivity difference between the two layers is considered. The resistivity of each layer is provided in section S6. The TaN layers are much more resistive than the CoFeB layer, and thus the difference between $J_{Total}$ and $J_{CoFeB}$ is larger than that of the Ta underlayer wires. Since $J_{CoFeB}$ becomes large for the TaN(Q=0.7%) underlayer with small $d$, we infer that the spin transfer torque dominates over the spin Hall torque in this regime.

## S6. Current induced effective field measurements

Current induced effective field is measured in the same manner as described in Ref. [1]. A Hall bar is patterned on the same wafer with the wires. To obtain the effective field, a sinusoidal constant amplitude voltage is applied to the Hall bar and the first and second harmonic Hall voltages are measured using lock-in amplifiers. The resistance does not change with the voltage within the range we apply, thus the excitation can be treated as a constant amplitude sinusoidal current. An in-plane magnetic field directed along or transverse to the current flow is applied to evaluate the longitudinal ($\Delta H_L$) and transverse ($\Delta H_T$) component of the effective field, respectively.

Contribution from the planar Hall effect[3] is taken into account in obtaining $\Delta H_L$ and $\Delta H_T$.

The underlayer thickness dependences of the anomalous ($\Delta R_{AHE}$) and planar ($\Delta R_{PHE}$) Hall effects are shown in Fig. S5(a-c); the ratio of $\Delta R_{PHE}$ to $\Delta R_{AHE}$ is plotted in Fig. S5(d-f). The planar Hall effect is less than 10% (in magnitude) of the anomalous Hall effect for all film structures.

Figure S6(a-f) show $\Delta H_L$ and $\Delta H_T$ normalized by the current flowing the underlayer as a function of the underlayer thickness. Since the resistivities of the CoFeB layer and the TaN underlayers are different, the current flow is not uniform. To estimate the magnitude of the spin Hall angle, the effective field per unit current density in the underlayer is calculated. The thick underlayer limit of $\Delta H_L$ and $\Delta H_T$ (i.e. the saturated values, defined as $\Delta H_L^{SAT}$ and $\Delta H_T^{SAT}$) shown in Fig. S6 is proportional to the spin Hall angle[4]. The size of $\Delta H_L^{SAT}$ and $\Delta H_T^{SAT}$ varies depending on the film structure, i.e. the size of $Q$. We infer that the boron concentration in the underlayer may partly play a role here by changing the scattering rate. Boron concentration in the underlayer can be inferred indirectly by the saturation magnetization value of the CoFeB layer, assuming that boron diffuses out from the CoFeB layer to the underlayer upon annealing. The saturation magnetization drops as the nitrogen concentration is increased, indicating that less boron is present in the underlayer as $Q$ is increased[2]. Interestingly, the variation of $\Delta H_T^{SAT}$ with $Q$ is similar to that of the interface magnetic anisotropy[2], which is primarily determined at the CoFeB|MgO interface. This indicates that the CoFeB|MgO interface may influence the size of $\Delta H_{T(L)}^{SAT}$. Further investigation is required to clarify the factors that determine the size of $\Delta H_L^{SAT}$ and $\Delta H_T^{SAT}$.

Resistivity of each film is evaluated using the Hall bars. The resistivity $\rho$ of the Ta(N) underlayers are $\rho \sim 191$, 375 and 908 $\mu\Omega \cdot cm$ for Q=0, 0.7 and 2.5%, respectively. The resistivity of the CoFeB layer is $\sim 111$ $\mu\Omega \cdot cm$[1].

## S7. One dimensional model of a domain wall

The one dimensional model[5] describing domain wall dynamics is used to understand the effect of the spin Hall effect, the DMI and the spin transfer torque. The dynamics of a domain wall is described by two parameters, its position $q$ and magnetization angle $\Psi$. For out of plane magnetized samples, the domain wall magnetization points along a direction within the film plane: we define $\Psi=0$ and $\pi$ corresponding to the Bloch wall and $\Psi=\pi/2$ and $3\pi/2$ as the Neel wall. The two coupled equations that describe the dynamics of $(q, \Psi)$ read:

$$(1+\alpha^2)\frac{\partial q}{\partial t} = \left[\frac{1}{2}\gamma\Delta H_K \sin 2\psi + \frac{\pi}{2}\gamma\Delta H_T \sin\psi - \frac{\pi}{2}\gamma\Delta(H_L + H_{DM})\cos\psi + u\right]$$
$$+ \alpha\left[\gamma\Delta Q H_Z + \gamma\Delta Q H_{SH}\sin\psi + \beta u\right] \quad (1a)$$

$$(1+\alpha^2)\frac{\partial \psi}{\partial t} = -\alpha\left[\frac{1}{2}\gamma H_K \sin 2\psi + \frac{\pi}{2}\gamma H_T \sin\psi - \frac{\pi}{2}\gamma(H_L + H_{DM})\cos\psi + \frac{u}{\Delta}\right]$$
$$+ \left[\gamma Q H_Z + \gamma Q H_{SH}\sin\psi + \beta\frac{u}{\Delta}\right] \quad (1b)$$

Here, $\gamma$ is the gyromagnetic ratio, $H_K$ is the magnetic anisotropy associated with the domain wall magnetization, $\Delta$ is the domain wall width and $\alpha$ is the Gilbert damping constant. Spin transfer torque is represented by $u = \frac{\mu_B P}{eM_S}J$, where $\mu_B$ and $e$ are the Bohr magnetron and the electron charge, $P$ and $M_S$ are the current spin polarization and saturation magnetization of the ferromagnetic material. $\beta$ is the non-adiabatic spin torque term[6]. $H_Z$, $H_T$ and $H_L$ correspond to the out of plane, in-plane transverse (transverse to the wire's long axis) and in-plane longitudinal (along the wire's long axis) fields, respectively. $Q$ represents the type of domain wall; $Q=+1$ for ↑↓ wall and $Q=-1$ for ↓↑ wall. The spin Hall effect is modeled[7] by an effective out of plane magnetic field $H_{SH}\sin\psi$ and the DMI is included as an offset in-plane field $H_{DM}$ directed along the wire's long axis[7-9]. For a domain wall spiral, the offset field $H_{DM}$ changes its sign depending on the type of domain wall ($Q$).

To describe experimental results using Eq. (1), we introduce the following parameter. Since $u$ and the spin Hall effective field scales with the current density, we define $u = -\tilde{P}j$ and $H_{SH} = -\tilde{\theta}_{SH} j$ [7]. Here $\tilde{P} \equiv -\frac{\mu_B P|J|}{|e|M_S}$, $\tilde{\theta}_{SH} \propto \theta_{SH} \frac{\pi}{2} \frac{\hbar|J|}{2eM_s t}$, $\theta_{SH}$ is the spin Hall angle, $t$ is the thickness of the magnetic layer and $j$ represent the direction of current, i.e. $j = \frac{J}{|J|}$. The sign of the spin Hall angle is set as the following: $\theta_{SH} > 0$ for Pt and $\theta_{SH} < 0$ for Ta. The chirality of the domain wall spiral is denoted by $S$: $S=1$ for right handed and $S=-1$ for left handed domain walls. The Dzyaloshinskii-Moriya offset field $H_{DM}$ can then be expressed using $Q$ and $S$ as: $H_{DM} = QS|H_{DM}|$. Substituting these parameters into Eq. (1a) and (1b) gives:

$$(1+\alpha^2)\frac{\partial q}{\partial t} = \left[\frac{1}{2}\gamma\Delta H_K \sin 2\psi + \frac{\pi}{2}\gamma\Delta H_T \sin\psi - \frac{\pi}{2}\gamma\Delta\left(H_L + QS|H_{DM}|\right)\cos\psi - \tilde{P}j\right]$$
$$+ \alpha\left[\gamma\Delta Q H_Z - \gamma\Delta Q \tilde{\theta}_{SH} j \sin\psi - \beta\tilde{P}j\right] \quad (2a)$$

$$(1+\alpha^2)\frac{\partial \psi}{\partial t} = -\alpha\left[\frac{1}{2}\gamma H_K \sin 2\psi + \frac{\pi}{2}\gamma H_T \sin\psi - \frac{\pi}{2}\gamma\left(H_L + QS|H_{DM}|\right)\cos\psi - \frac{\tilde{P}j}{\Delta}\right]$$
$$+ \left[\gamma Q H_Z - \gamma Q \tilde{\theta}_{SH} j \sin\psi - \beta\frac{\tilde{P}j}{\Delta}\right] \quad (2b)$$

Left handed walls: ↑←↓ wall: $S=-1$, $Q=1$, ↓→↑ wall: $S=-1$, $Q=-1$

Right handed walls: ↑→↓ wall: $S=1$, $Q=1$, ↓←↑ wall: $S=1$, $Q=-1$

$\tilde{P} > 0$ for positively spin polarized materials (e.g. Py, Co, CoFeB)

$\tilde{\theta}_{SH} > 0$ for Pt, Pd, etc., $\tilde{\theta}_{SH} < 0$ for Ta, W, etc.

$j=+1$ for current flowing along +x, $j=-1$ for current flowing along –x.

When $\Psi$ is small, Eqs. (2a) and (2b) can be linearized to calculate the domain wall velocity (below the Walker breakdown limit). The solution is given as:

$$v_{DW} \equiv \frac{\partial q}{\partial t} = \frac{\pm\gamma\Delta\frac{\pi}{2}Q\tilde{\theta}_{SH}j}{-Q\tilde{\theta}_{SH}j + \alpha\left(\mp H_K - \frac{\pi}{2}H_T\right)}\left(H_L + QS|H_{DM}|\right) \qquad (3)$$

$$-\frac{-Q\tilde{\theta}_{SH}j + \beta\left(\mp H_K - \frac{\pi}{2}H_T\right)}{-Q\tilde{\theta}_{SH}j + \alpha\left(\mp H_K - \frac{\pi}{2}H_T\right)}\tilde{P}J + \gamma\Delta\frac{\left(\mp H_K - \frac{\pi}{2}H_T\right)}{-Q\tilde{\theta}_{SH}j + \alpha\left(\mp H_K - \frac{\pi}{2}H_T\right)}QH_Z$$

where the upper and lower (plus/minus) sign indicate cases for $\Psi$ close to 0 and $\pi$, respectively.

The domain wall velocity is calculated numerically and plotted in Fig. S7(a-d) for cases with (bottom panels) and without (top panels) contribution from the spin transfer torque (STT). The spin Hall effective field is determined by the longitudinal field $\Delta H_L$ (Fig. 3), and is illustrated by the red big arrow in Fig. S7(a-d) when a current (+I) is applied as indicated by the blue large arrow. The direction to which a domain wall moves (for +I) upon application of spin Hall torque (STT) is shown by the red (orange) thin arrow. The velocity is calculated for each wall type ( ↑ ↓ and ↓ ↑ walls) and each chirality (left or right).

The calculations show that for left handed walls, the STT increases the magnitude of the wall velocity at zero field and the magnitude of the compensation field $H_L^*$. In contrast, for right handed walls, the zero field velocity and $H_L^*$ both decreases in magnitude when STT is added. The compensation field can change its sign, as shown in Fig. S7(c,d), when the STT contribution becomes larger than that of the spin Hall torque. Note that the relation between the wall handedness and the STT effect will be reversed when the direction of the spin Hall effective field is altered, for example, in systems with Pt underlayers.

**References**

1. Kim, J., Sinha, J., Hayashi, M., Yamanouchi, M., Fukami, S., Suzuki, T., Mitani, S. & Ohno, H. Layer Thickness Dependence of the Current Induced Effective Field Vector in Ta|Cofeb|Mgo. *Nat. Mater.* **12**, 240 (2013).
2. Sinha, J., Hayashi, M., Kellock, A. J., Fukami, S., Yamanouchi, M., Sato, M., Ikeda, S., Mitani, S., Yang, S. H., Parkin, S. S. P. & Ohno, H. Enhanced Interface Perpendicular Magnetic Anisotropy in Ta|Cofeb|Mgo Using Nitrogen Doped Ta Underlayers. *Appl. Phys. Lett.* **102**, 242405 (2013).
3. Garello, K., Mihai Miron, I., Onur Avci, C., Freimuth, F., Mokrousov, Y., Blügel, S., Auffret, S., Boulle, O., Gaudin, G. & Gambardella, P. Symmetry and Magnitude of Spin-Orbit Torques in Ferromagnetic Heterostructures. arXiv:1301.3573 (2013).
4. Liu, L., Pai, C.-F., Li, Y., Tseng, H. W., Ralph, D. C. & Buhrman, R. A. Spin-Torque Switching with the Giant Spin Hall Effect of Tantalum. *Science* **336**, 555 (2012).
5. Malozemoff, A. P. & Slonczewski, J. C. *Magnetic Domain Walls in Bubble Material*. (Academic Press, 1979).
6. Thiaville, A., Nakatani, Y., Miltat, J. & Suzuki, Y. Micromagnetic Understanding of Current-Driven Domain Wall Motion in Patterned Nanowires. *Europhys. Lett.* **69**, 990 (2005).
7. Thiaville, A., Rohart, S., Jue, E., Cros, V. & Fert, A. Dynamics of Dzyaloshinskii Domain Walls in Ultrathin Magnetic Films. *Europhys. Lett.* **100**, 57002 (2012).
8. Ryu, K.-S., Thomas, L., Yang, S.-H. & Parkin, S. Chiral Spin Torque at Magnetic Domain Walls. *Nat. Nanotechnol.* **8**, 527 (2013).
9. Emori, S., Bauer, U., Ahn, S.-M., Martinez, E. & Beach, G. S. D. Current-Driven Dynamics of Chiral Ferromagnetic Domain Walls. *Nat Mater* **12**, 611 (2013).

**Figure captions**

**Figure S1. Magnetic properties of Ta(N)|CoFeB|MgO heterostructures.** The saturated magnetic moment per unit volume ($M/V$), anisotropy field ($H_K$) and the magnetic anisotropy energy ($K_{EFF}$) are plotted as a function of the underlayer thickness in (a-c) for films with different underlayers.

**Figure S2: Magneto-optical properties of the heterostrucutres**. (a) Out of plane hysteresis loops measured using the Kerr microscopy for wires with TaN(Q=0.7%) underlayer; loops with various underlayer thicknesses are shown. The y-axis indicates the average CCD intensity $I$ of the wire. Each plot is shifted vertically for clarity. (b) Change in the CCD intensity with the out of plane field $\Delta I/I_0$ is plotted as a function of the underlayer thickness for films with different underlayers. Positive (negative) $\Delta I/I_0$ represents bright (dark) contrast for $+M_z$.

**Figure S3. Domain wall propagation field**. (a,b) Exemplary Kerr images showing domain wall propagation with out of plane field $H_Z$. The wire is composed of 3.6 nm thick TaN(Q=0.7%) underlayer. As $H_Z$ is increased in magnitude, domain walls move to expand the energetically favorable domains. (a) Positive and (b) negative out of plane field is applied to move the walls. Values of $H_Z$ are indicated in each panel. Images are taken at a rate of ~1 Hz. (c-e) Domain wall propagation field $H_P$ plotted as a function of the underlayer thickness for films with different underlayers.

**Figure S4. Current density required for domain wall motion.** The underlayer thickness dependence of the current density $J$ needed for moving a domain wall are plotted in (a-c) for films with different underlayers. *J* corresponds to the maximum pulse amplitude used for each wire as shown in Fig. 2. The solid symbols correspond to *J* if one assumes a uniform current flow across the two layers (Ta(N) and CoFeB). The open symbols show *J* for the current that flows through the CoFeB layer when the resistivity difference between the two layers is considered.

**Figure S5. Anomalous and planar Hall effect.** The underlayer thickness dependences of the anomalous and planar Hall effects are plotted in (a-c). The anomalous ($\Delta R_{AHE}$) and planar ($\Delta R_{PHE}$) Hall resistances are obtained by measuring the change in the Hall resistance when the field is swept along the film normal or rotated within the film plane, respectively. (d-f) Ratio of the planar Hall to anomalous Hall resistances $\Delta R_{PHE}/\Delta R_{AHE}$ is plotted as a function of the underlayer thicknesses for films with different underlayers.

**Figure S6. Current induced effective fields.** The transverse ($\Delta H_T$) and longitudinal ($\Delta H_L$) components of the current induced effective field are plotted as a function of the underlayer thickness in (a-c) and (d-f), respectively, for films with different underlayers. The effective field is normalized by the current density that flows into the underlayer. Since the resistivity of the CoFeB layer and the underlayer is different, such normalization gives different results from that shown in Fig. 3. Here, we provided the effective field if we were to apply $1\times10^8$ A/cm$^2$ of current density to the underlayer.

**Figure S7. Numerical calculations using the one dimensional model of a domain wall.**

(a-d) Numerically calculated domain wall velocity plotted as a function of an in-plane longitudinal field $H_L$ (along the current flow) for left handed (S=−1) ↓↑ (a), ↑↓ (b) walls and right handed (S=+1) ↓↑ (c), ↑↓ (d) walls. Calculation results are shown for cases with (bottom panel) and without (top panel) the spin transfer torque. Squares and circles show numerical calculations for +I ($j$=+1) and –I ($j$=−1), respectively. The solid lines indicate analytical solutions (Eq. (3)).

The black thick arrows in each cartoon show the magnetization direction including that of a domain wall. The yellow spheres illustrate the electrons; the arrow penetrating each electron represents its spin direction. Motion of electrons caused by the spin Hall effect (in Ta(N)) are illustrated. The red large arrow indicates the spin Hall effective field $\Delta H_L$ when current (+I) is applied. The direction to which a domain wall moves with spin Hall torque and spin transfer torque are indicated by the red and orange thin arrows, respectively, placed next to each cartoon.

Values used in the calculations are: $H_K$=200 Oe, $\Delta$=10 nm, $\alpha$=0.05, $\beta$=0, $|H_{DM}|$=50 Oe, $\tilde{\theta}_{SH}$=−10 Oe, $\tilde{P}$=0 m/s (top panels) and 20 m/s (bottom panels). (a) Q=−1, S=−1, (b) Q=+1, S=−1, (c) Q=−1, S=+1, (d) Q=+1, S=+1.

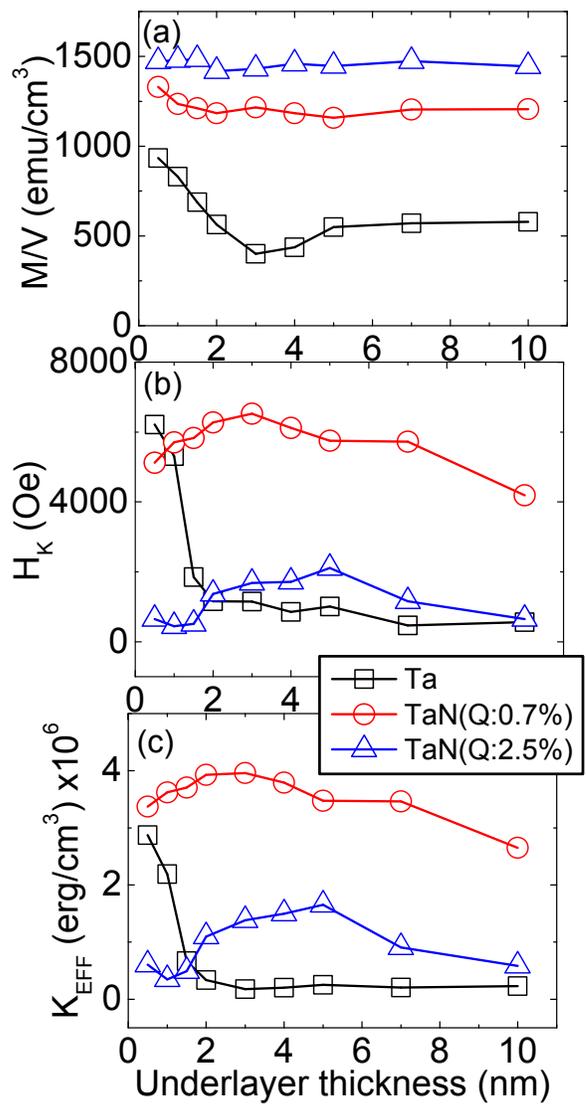

Fig. S1

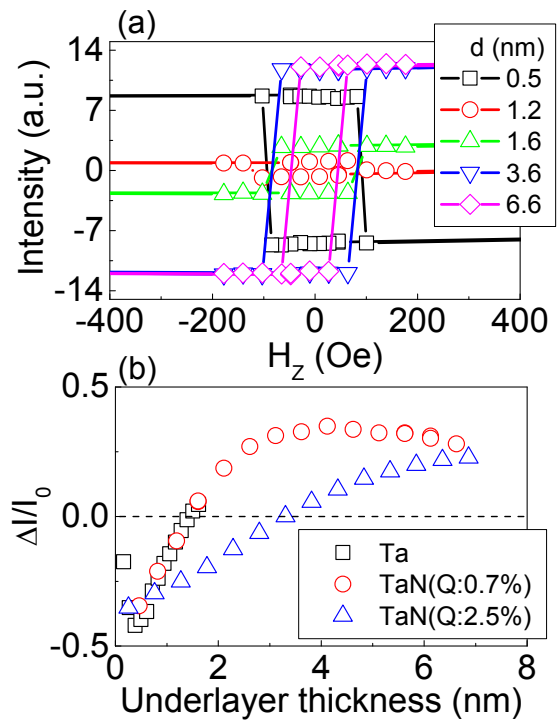

Fig. S2

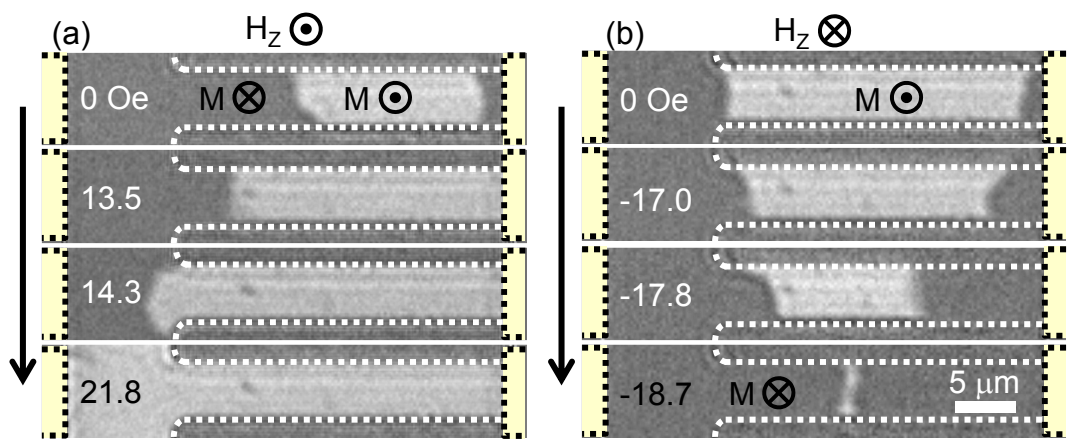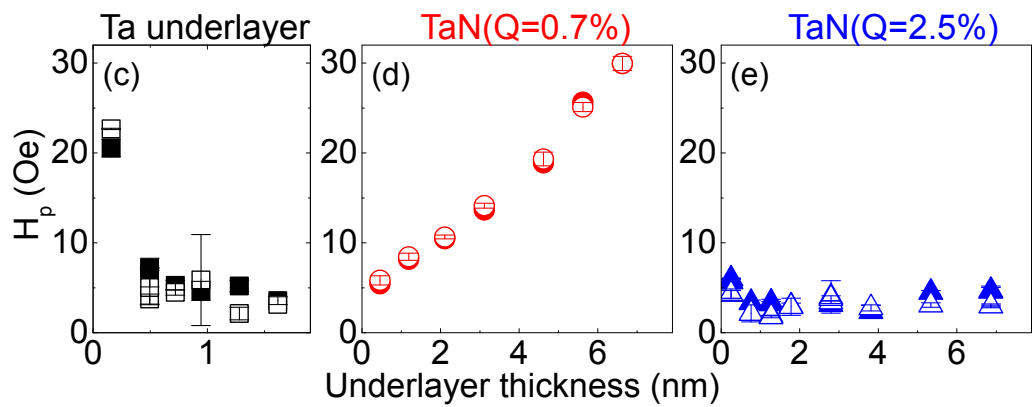

Fig. S3

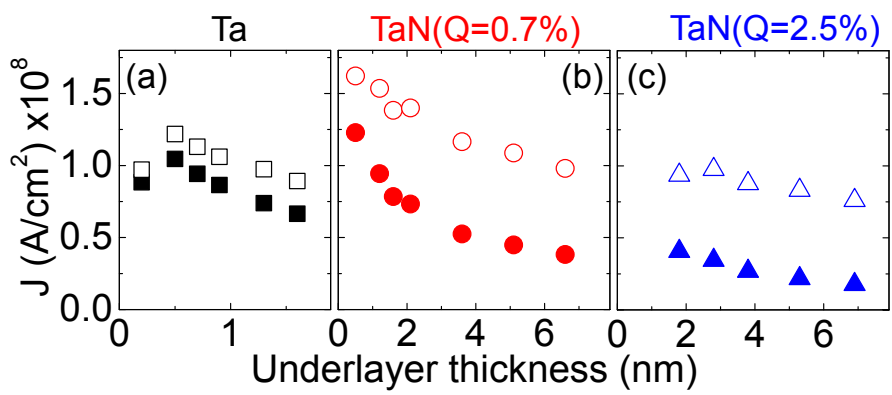

Fig. S4

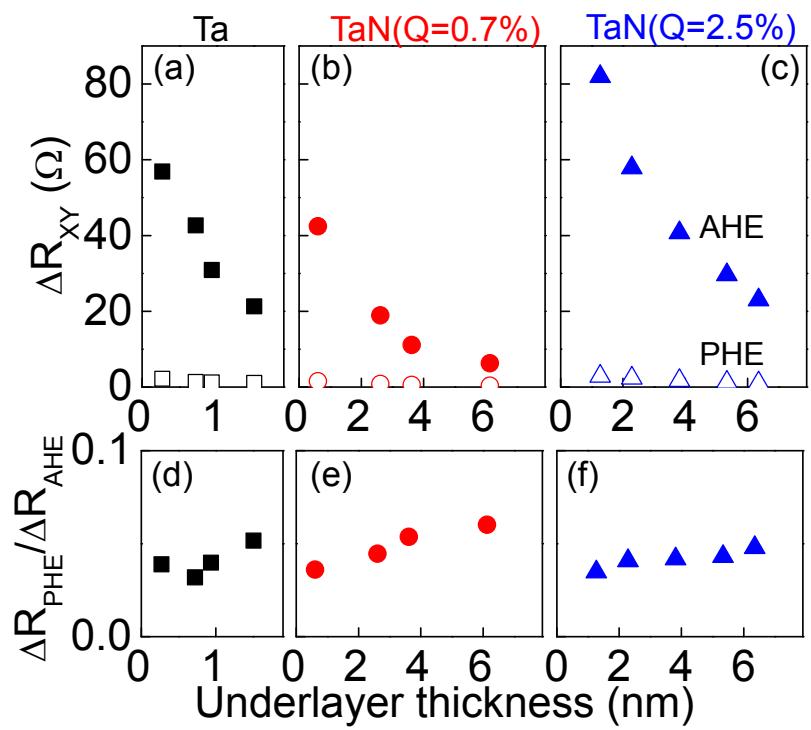

Fig. S5

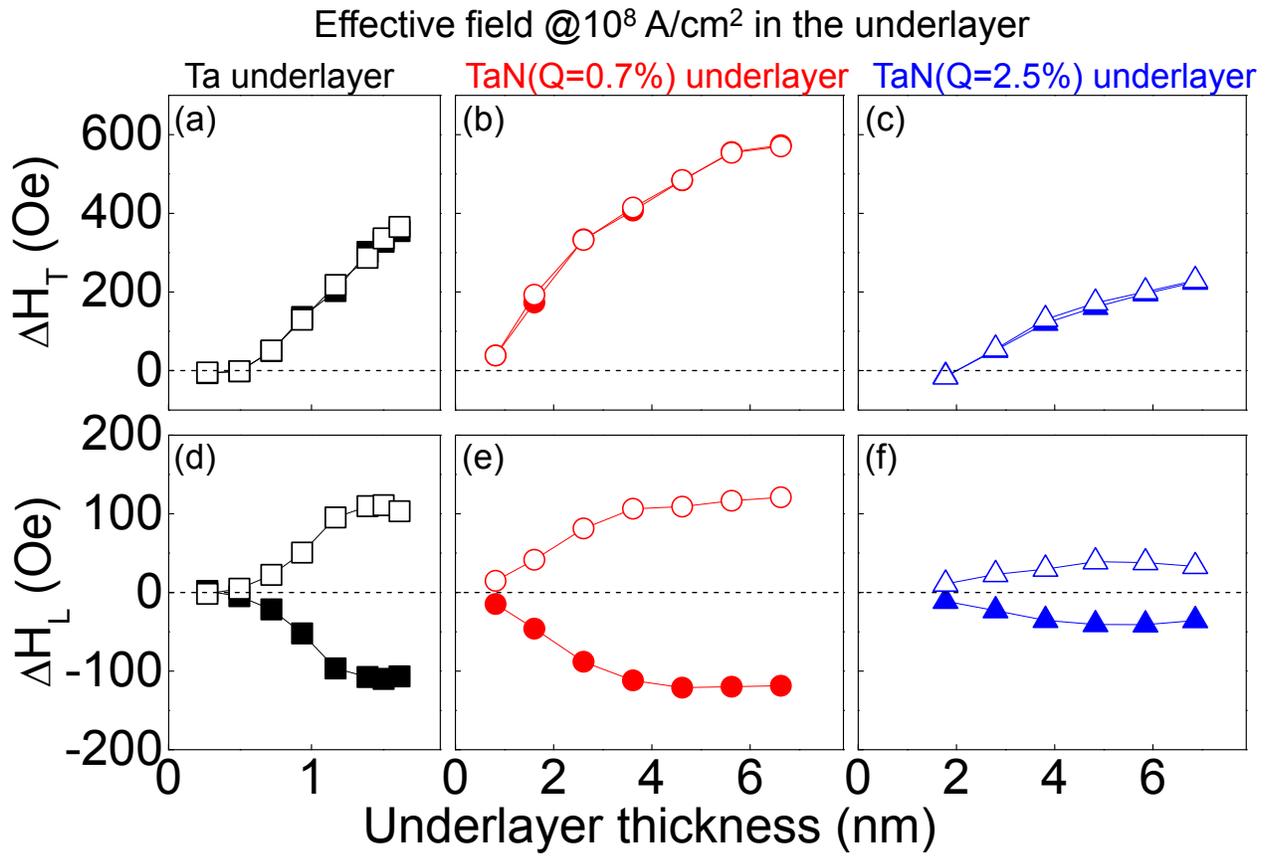

Fig. S6

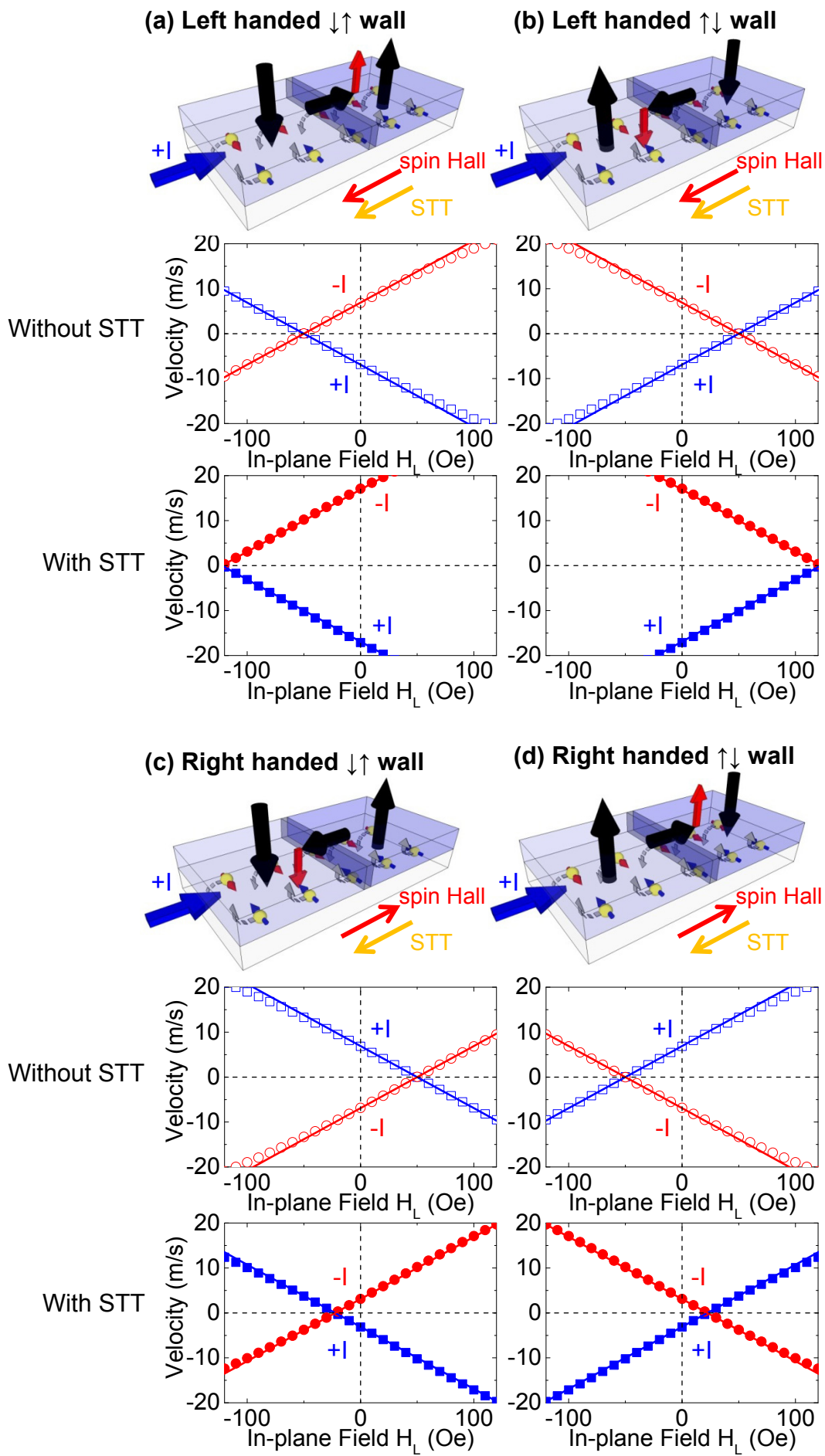

Fig. S7